\documentclass[fleqn,usenatbib,useAMS]{mnras}

\usepackage[T1]{fontenc}

\usepackage{graphicx}	
\usepackage{amsmath}	

\usepackage{amssymb}	
\usepackage{multirow,booktabs}
\usepackage{xcolor}
\usepackage{ulem}
\usepackage{natbib}
\usepackage{ae,aecompl}
\usepackage{bm}

\newcommand{\kms}{\mathrm{\ km\ s^{-1}}} 

\newcommand{\au}{\mathrm{\ au}}
\newcommand{\pc}{\mathrm{\ pc}}

\newcommand{\days}{\mathrm{\ d}}
\newcommand{\yr}{\mathrm{\ yr}}
\newcommand{\myr}{\mathrm{\ Myr}}

\newcommand{\mx}{\mathrm{max}}

\newcommand{\out}{\mathrm{out}}
\newcommand{\inner}{\mathrm{in}}

\title[Partial TDEs]{Partial Tidal Disruption Events by Intermediate-mass Black Holes in  Supermassive and Intermediate-mass Black Hole Binaries}

\author[X.-J. Wu, Y.-F. Yuan, Y Luo and W Lin ]{
        Xiao-Jun Wu,$^{1,2}$ 
Ye-Fei Yuan,$^{3,4}$\thanks{Corresponding author. E-mail: yfyuan@ustc.edu.cn} 
Yan Luo,$^{3,4}$ 
and Wenbin Lin$^{1,5}$\thanks{Corresponding author. E-mail: lwb@usc.edu.cn}
\\
$^{1}$School of Mathematics and Physics, University of South China, Hengyang, 421001, China\\
$^{2}$School of Nuclear Science and Technology, University of South China, Hengyang 421001, China\\
$^{3}$School of Astronomy and Space Science, University of Science and Technology of China, Hefei 230026, China\\
$^{4}$CAS Key Laboratory for Research in Galaxies and Cosmology, University of Science and Technology of China, Hefei 230026, China\\
$^{5}$School of Physical Science and Technology, Southwest Jiaotong University, Chengdu, 610031, China\\
\\
}

\date{Accepted XXX. Received YYY; in original form ZZZ}

\pubyear{2024}

\graphicspath{{./}{figures/}}
\begin{document}
\label{firstpage}
\pagerange{\pageref{firstpage}--\pageref{lastpage}}
\maketitle

\begin{abstract}
In the centers of galaxies, stars that orbit supermassive black hole binaries (SMBHBs) can undergo tidal disruptions due to the Lidov-Kozai mechanism. Nevertheless, most previous researches have predominantly focused on full tidal disruption events (FTDEs). In this study, we employ N-body simulations to investigate partial tidal disruption events (PTDEs) induced by intermediate-mass black holes (IMBHs) in SMBH-IMBH binaries, taking into account consideration the IMBH's mass, semi-major axis, and eccentricity of the outer orbit.
Our findings indicate that, in comparison to FTDEs, the majority of tidal disruption events are actually PTDEs. Furthermore, we find that a significant number of stars experiencing partial disruption ultimately get captured by the IMBH, potentially leading to repeating flares. By comparing the period of the periodic eruptions observed in ASASSN-14ko, we find that PTDEs in a specific SMBH-IMBH binary system can align with the observed period if the SMBH has a mass of $10^7\rm{\ M_\odot}$, the IMBH has a mass smaller than approximately $10^5\rm{\ M_\odot}$, the eccentricity of the SMBH-IMBH binary exceeds approximately $0.5$, and the semi-major axis of the SMBH-IMBH binary is larger than approximately $0.001\pc$. 
Moreover, our model effectively accounts for  the observed period derivative for ASASSN-14ko ($\dot{P}=-0.0026\pm 0.0006$), and  
our results also imply that some quasi-periodic eruptions may be attributed to PTDEs occurring around SMBH-IMBH binaries.
\end{abstract}

\begin{keywords}
stars: kinematics and dynamics -- Galaxy: centre -- Galaxy: kinematics and dynamics -- transients: tidal disruption events -- methods: numerical
\end{keywords}



\section{Introduction} \label{sec:intro}
When stars approach too close towards a supermassive black hole (SMBH), they could be tidally disrupted by the SMBH, giving rise to multiband flares from X-rays to radio bands \cite[e.g.,][]{bade1996,komossa1999,esquej2008,komossa2008,bloom2011,vanVelzen2011,
vanVelzen2016,cenko2012,anderson2020,goodwin2023}. This phenomenon is known as a tidal disruption event (TDE) \citep{rees1988}. If a star is completely destroyed during the process, it is referred to as a full TDE (FTDE), while if only a partial destruction occurs, leaving behind a remnant, it is termed a partial TDE (PTDE). For PTDEs, the remnants can acquire a kick velocity due to the asymmetry in mass loss during the disruption process \citep{manukian2013}. If the orbital energy of the remnants is insufficient to overcome the gravitational potential of the SMBH, they will orbit the SMBH in a bound trajectory and may undergo subsequent PTDEs or FTDEs.

There are several candidates that have been proposed as potential results of PTDEs, including HLX-1 \citep{lasota2011}, IC 3599 \citep{campana2015,grupe2015}, AT 2018hyz \citep{gomez2020}, AT2018fyk \citep{wevers2022b}, ASASSN-14ko \citep{payne2021}, Swift J023017.0+283603 \citep{evans2023,guolo2024}, eRASSt J045650.3-203750 \citep{liu2023}, and X-ray quasi-periodic eruptions (QPEs) \cite[e.g.][]{miniutti2019,giustini2020,song2020,arcodia2021,chakraborty2021}.

In order to generate repeating PTDEs, stars need to be captured into eccentric orbits. Traditional two-body relaxation \cite[e.g.,][]{alexander2017} can bring stars into close orbits around a SMBH, but the long orbital periods of the captured stars do not align with the short periods observed in some systems, such as ASASSN-14ko, in which repeating eruptions with a period of $\sim115\days$ are reported by \citet{payne2021}. \citet{cufari2022} showed that the orbital period generated by two-body relaxation is $\sim1000$ times larger than the period of ASASSN-14ko. Consequently, it seems necessary for an alternative mechanism to drive the partially disrupted star onto its $\sim115\days$ orbit about the SMBH.

As demonstrated by \citet{cufari2022}, the Hills mechanism \citep{hills1988} can account for the repetitive flares observed in ASASSN-14ko. According to this mechanism, a binary star is disrupted by the tidal force exerted by the SMBH, resulting in a captured star on an eccentric orbit around the SMBH, as well as an ejected star. The captured star undergoes partial disruption at each pericenter passage of its orbit.

The post-merger signatures observed in the host galaxy ESO 253-G003 of ASASSN-14ko \citep{tucker2021,payne2022a} suggest the presence of a SMBH binary (SMBHB) at the galactic center, which could give rise to PTDEs. Previous researches have extensively investigated full tidal disruptions of stars in SMBHBs \cite[e.g.,][]{chen2009,chen2011,wegg2011,fragione2018,wu2018,li2019,mockler2023}. Among these studies, one leading model is hierarchical systems, in which stars form the inner binary with one component SMBH, while the outer SMBH and the center of mass of the inner binary constitute the outer binary. When the inclination angle between the orbital planes of the inner and outer binaries falls within the range of $40^\circ\lesssim i\lesssim 140^\circ$, the Lidov-Kozai (LK) mechanism \citep{lidov1962,kozai1962,naoz2016} can drive the eccentricity of the inner binary to $\sim 1$. Consequently, both full and partial TDEs are anticipated. 

Recently, \citet{melchor2023} investigated the combined effects of the LK mechanism and two-body relaxation on TDEs in SMBHBs. Their findings revealed that relying solely on the LK mechanism is insufficient to induce significant changes in the semi-major axis of inner binaries. Instead, the introduction of two-body relaxation contributes to diffusive alterations in the energy and angular momentum of stars. This process has the potential to transfer stars into close orbits around the secondary SMBH, leading to the occurrence of PTDEs with periods shorter than or equal to $30\ \rm years$. Notably, their work implies that certain recurring flares, exemplified by events like ASASSN-14ko, find plausible explanations in PTDEs within systems featuring SMBHBs.

In this study, we investigate the potential occurrence of PTDEs in systems involving SMBHBs by means of N-body simulations. Specifically, we consider a configuration where the SMBHB consists of an intermediate-mass black hole (IMBH) and a SMBH, and stars orbit around the inner IMBH, forming the inner binary, and the SMBH forms the outer binary with the center of mass of the inner binary. Through high-precision N-body simulations, we examine the likelihood of stars undergoing partial disruptions by the IMBH. We explore the influence of the IMBH mass, as well as the semi-major axis and eccentricity of both the inner and outer binaries. Our aim is to compare our results with the period observed in ASASSN-14ko \citep{payne2021,payne2022a,payne2023}.

This paper is structured as follows. In Section \ref{sec:kl}, we provide a brief overview of the Lidov-Kozai mechanism. Section \ref{sec:ptde} provides a brief introduction to PTDEs. The methodology and initial conditions are described in Section \ref{sec:initial}. The results are presented in Section \ref{sec:results}. Lastly, in Section \ref{sec:conclusions_and_discussions}, we summarize our findings and engage in discussions.

\section{Lidov-Kozai mechanism} \label{sec:kl}

\begin{figure}
	\centering
	\includegraphics[width=0.88\columnwidth]{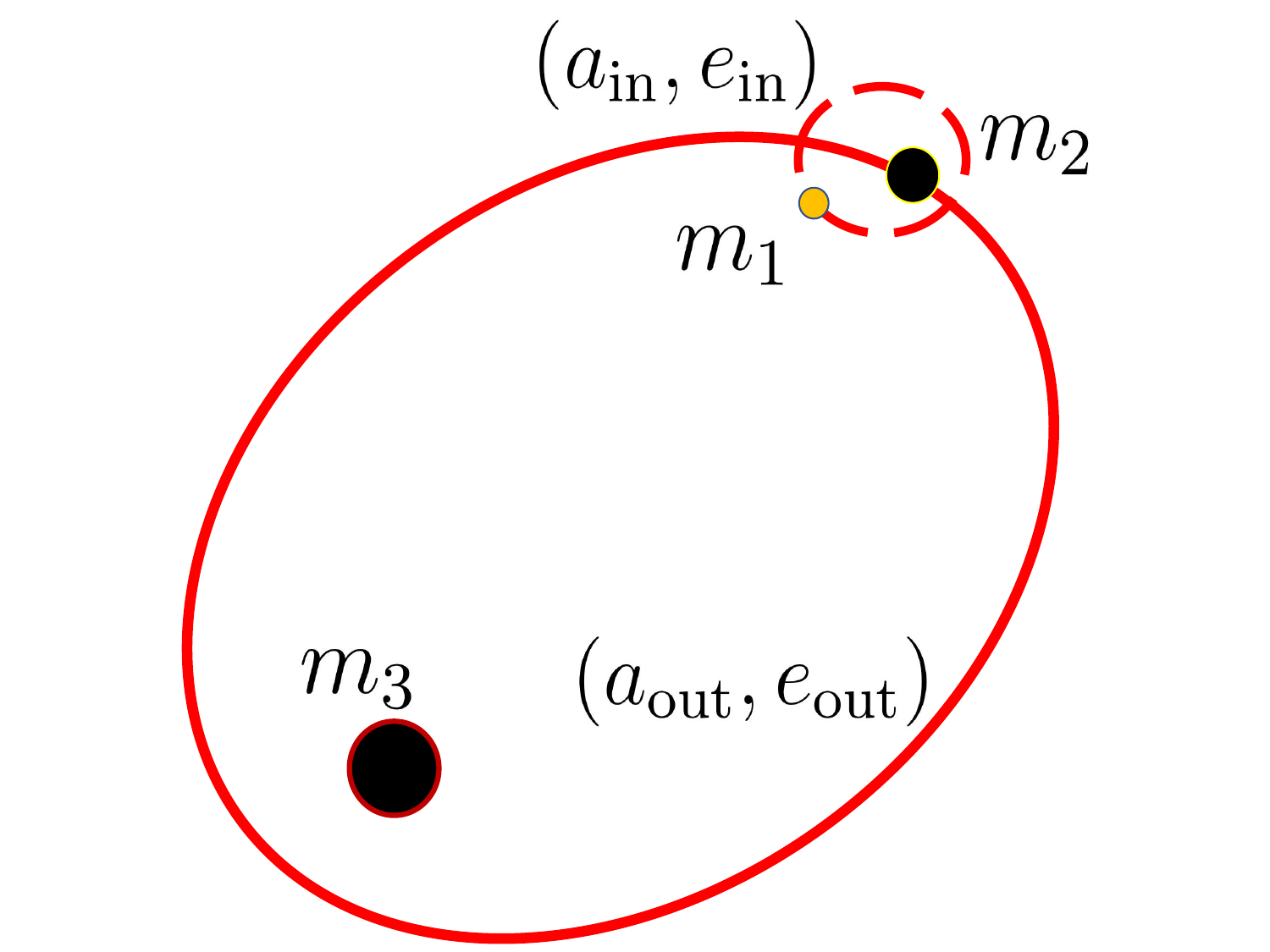}
	\caption{Schematic illustration of the model, which consists of the inner binary (which is composed of a star and an IMBH) and
	the outer binary (which is composed of a SMBH and the center of mass of the inner binary). The semi-major axis and eccentricity of the inner and outer binaries are $(a_\inner, e_\inner)$ and $(a_\out, e_\out)$ respectively. The masses of the star, IMBH and SMBH are $m_1$, $m_2$ and $m_3$, respectively.}
	\label{fig:model}
\end{figure}

As is shown in Fig. \ref{fig:model}, we assume the model consists of the inner binary and the outer binary. The semi-major axis and eccentricity are $a$ and $e$, with the subscripts ``\textit{\rm in}'' and ``\textit{\rm out}'' denoting the inner and outer binaries, respectively. The masses of the star, IMBH and the SMBH are $m_1$, $m_2$ and $m_3$, respectively.

Because of the interaction between the inner and outer binaries, the eccentricity of the inner binary and inclination angle between two binaries will oscillate, which is called the Lidov-Kozai mechanism \citep{lidov1962,kozai1962}. At the quadruple level (i.e. the second order in the semi-major axial ratio of the inner and outer orbits) of approximation, for the circular outer orbit (i.e. $e_\out=0$), with the inclination angle $i$ is in the range of $40^\circ\lesssim i \lesssim140^\circ$, the eccentricity will oscillations on a timescale of \citep{naoz2016}
\begin{equation}\label{eq:tlk}
	t_{\rm k} = \frac{8}{15\pi}\frac{P_\out^2}{P_\inner}(1-e_\out^2)^{3/2},
\end{equation}
where $P_\inner$ and $P_\out$ are periods of the inner and outer binaries, respectively.
The maximum eccentricity that the inner binary can reach is about
\begin{equation}
	e_\mx = \sqrt{1-\frac{5}{3}(\cos{i})^2}.
\end{equation}

When the outer orbit is eccentric (i.e., $e_\out>0$), the eccentricity oscillation of the inner binary can exhibit significantly larger amplitude, leading to orbital flips \citep{lithwick2011} and even direct collisions \citep{katz2012}. These phenomena arise due to resonances from higher-order harmonics of the octupole-level Hamiltonian, known as the eccentric Lidov-Kozai mechanism \citep{naoz2016}.
The corresponding timescale is given by \citep{li2015}
\begin{equation}
t_{\rm oct}=\frac{t_{\rm k}}{\epsilon},
\end{equation}
where
\begin{equation}
\epsilon\equiv \frac{a_\inner}{a_\out}\frac{e_\out}{1-e_\out^2}<0.1,
\end{equation}
indicating that the triple system is hierarchical.

\section{Partial tidal disruptions} \label{sec:ptde}

According to the classical theory (e.g. \cite{rees1988}; \cite{evans1989}), a star can undergo a tidal disruption when it crosses the tidal sphere of a black hole, which is estimated by the following expression:
\begin{equation}
r_{\rm t} = \left(\frac{M_{\rm bh}}{m_\ast}\right)^{1/3}R_\ast = \beta r_{\rm p},
\end{equation}
where $M_{\rm bh}$ is the mass of the black hole, $m_\ast$ and $R_\ast$ are the mass and radius of the star, respectively, $\beta\equiv r_{\rm t}/r_{\rm p}$ is the penetration factor, and $r_{\rm p}$ represents the periapsis of the star. However, hydrodynamical simulations \citep{guillochon2013,mainetti2017} have shown that the tidal force from the black hole can only overcome the surface gravity of the star when $r_{\rm p}\sim r_{\rm t}$.

In fact, there exists a critical parameter $\beta_{\rm d}$, which determines the threshold for full disruption. When the penetration factor follows $\beta \ge \beta_{\rm d}$ \cite[e.g. ][]{guillochon2013,mainetti2017}, the star can be fully disrupted. Here, $\beta_{\rm d}$ depends on the structure of the star. For the polytropic index $\gamma=4/3\ (5/3)$, $\beta_{\rm d}$ is $1.85\ (0.9)$ \citep{guillochon2013}. When $\beta_{\rm p} \le \beta \le \beta_{\rm d}$, the star can only be partially disrupted by the black hole. We adopt a lower limit of $\beta_{\rm p}=0.6$ based on the work by \citet{guillochon2013}.

For a partially disrupted star, the outer parts of the star are stripped by the tidal field of the black hole, giving rise to tidal disruption flares \citep{strubbe2009}. If the remnant falls back towards the black hole again, a repeating flare emerges \cite[e.g.][]{zhong2022}. The mass loss by the remnant during each pericenter passage is given by \citep{guillochon2013}
\begin{equation} \label{eq:mass_loss}
	\Delta m = C_\gamma m_\ast,
\end{equation}
where $m_\ast$ is the mass of the remnant of the star before each disruption. If the polytropic index is assumed to be $\gamma=4/3$, then we have
\begin{equation}
	C_{4/3} = \rm{exp}\Big[\frac{12.996-31.149\beta+12.865\beta^2}{1-5.3232\beta+6.4262\beta^2}\Big].
\end{equation}

The asymmetry in the mass loss near the inner and outer Lagrange points L1 and L2 induces a kick velocity imparted to the remnant \citep{liu2013,manukian2013}. In this work, the kick velocity is estimated by \citep{manukian2013}
\begin{equation}\label{eq:vkick}
v_{\rm kick}=(0.0745+0.0571\beta^{4.539})v_{\rm esc},
\end{equation}
where $v_{\rm esc}$ is the escape velocity.  
As the specific orbital angular momentum of the remnant is nearly unchanged after a PTDE
\citep{ryu2020}, the $v_{\rm kick}$ is added to the radial direction of the remnant \citep{zhong2022}.
When the kick velocity $v_{\rm kick}$ is large enough, the remnant could be ejected away from the black hole.
Notice that the orbital energy is the sum of the kinetic energy and potential energy. Therefore, 
this happens if the orbital energy between the black hole and the remnant is positive. If the orbital energy is negative, meaning that the kinetic energy of the remnant cannot overcome the potential energy between the remnant and the black hole, the remnant will be bound to the black hole and could experience a second disruption.

\section{Method}\label{sec:initial}
As depicted in Fig. \ref{fig:model}, our model is a triple system comprising inner and outer binaries. The inner binary consists of a star and an IMBH, while the outer binary comprises a SMBH and the center of mass of the inner binary. In our triple systems, we do not consider two-body interactions arising from the population of stars in galactic centers, which have been explored by \citet{melchor2023}. 

\subsection{Initial condition}

The initial conditions are presented in Table \ref{tab:initial_conditions}. In that table, for the SMBH, we set the mass as $m_3=10^{7}\rm\ M_\odot$, similar to the SMBH in ASASSN-14ko \citep{payne2021}. The mass of the IMBH is assumed to be $m_2=10^3, 10^4$, and $10^5\rm\ M_\odot$. 
The semi-major axis and the eccentricity of the outer binary are chosen as $a_\out=0.01, 0.001\pc$ and $e_\out=0, 0.5, 0.7$, respectively. 
Regarding the inner binary, we assume the star has an initial mass of $m_1=1\rm\ M_\odot$, a radius of $R_\ast=1\ R_\odot$, and a polytropic index of $\gamma=4/3$.
The semi-major axis of the inner binary (i.e., $a_\inner$) is randomly selected within the Hill sphere of the IMBH at its orbital pericenter with respect to the SMBH \cite[e.g.][]{fragione2018},
\begin{equation}
r_{\rm hill} = a_\out(1-e_\out)\left(\frac{m_2}{m_3}\right)^{1/3}.
\end{equation}
The eccentricity of the inner binary (i.e., $e_\inner$) is randomly chosen within $(0,1)$.

It's worth noting that our choice of $a_\out$ is smaller than that in \citet{fragione2018}, who considered $0.01\pc \leq a_\out \leq 0.1\pc$. We have two primary considerations for this choice. Firstly, the SMBH-IMBH binary might experience a stall when $a_\out \sim a_{\rm h} \approx Gm_2/(4\sigma^2)$ \citep{merritt2006}, where $\sigma$ represents the stellar velocity dispersion, and $a_{\rm h}$ is the hard-binary separation. In our calculations, assuming $m_1=10^7\ M_\odot$ and $\sigma\approx 100\kms$ \cite[e.g.,][]{gultekin2009}, we find that $a_{\rm h} \approx 0.01\pc(m_2/10^5M_\odot) \lesssim 0.01\pc$ for $m_2 \le 10^5\ M_\odot$. Secondly, the SMBH-IMBH binary could merge within $3-4\rm\ Gyr$ if the galaxy hosts a nuclear star cluster \citep{arca-sedda2018}. Therefore, the semi-major axis of the SMBH-IMBH binary could be even smaller than the hard-binary separation.

\begin{table}
	\caption{Model parameters (from the second column to the last column): the mass of the SMBH ($m_3$), the mass of the IMBH ($m_2$), the semi-major axis  ($a_{\rm out}$) and the eccentricity of the outer binary ($e_{\rm out}$), and the semi-major axis and eccentricity of the inner binary, which are randomly selected from $(0,r_{\rm hill})$ and $(0,1)$, respectively.}
	\centering
	\begin{tabular}{c c c c c c c}
	\hline\hline
	Model & $m_3/{\rm M_\odot}$ & $m_2/{\rm M_\odot}$ & $a_{\rm out}/{\rm pc}$  & $e_{\rm out}$ & $a_\inner$ & $e_\inner$
	\\[0.5ex]
	\hline
	M01 & $10^7$ & $10^3$ & 0.01 & 0.5 & $(0,r_{\rm hill})$ & $(0,1)$\\
	M02 & $10^7$ & $10^3$ & 0.001 & 0.5 & $(0,r_{\rm hill})$ & $(0,1)$\\
	M03 & $10^7$ & $10^4$ & 0.01 & 0 & $(0,r_{\rm hill})$ & $(0,1)$\\
	M04 & $10^7$ & $10^4$ & 0.01 & 0.5 & $(0,r_{\rm hill})$ & $(0,1)$\\
	M05 & $10^7$ & $10^4$ & 0.01 & 0.7 & $(0,r_{\rm hill})$ & $(0,1)$\\
	M06 & $10^7$ & $10^5$ & 0.01 & 0.5 & $(0,r_{\rm hill})$ & $(0,1)$\\
	\hline 
	\end{tabular}
	\label{tab:initial_conditions}
\end{table}

The inclination angle between the orbital planes of the inner and outer binaries is uniformly sampled from $\cos{i}\in [-1,1]$.
The remaining initial conditions, including the longitudes of the ascending nodes and the arguments of periapsis of the two binaries ($\omega$ and $\Omega$, respectively), are uniformly chosen from the range of 0 to $2\pi$. The origin of the frame of reference is set at the center of mass of the system, with the outer binary serving as the reference plane (X-Y plane).

\subsection{The lifetime of the systems}

In this section, we will estimate the lifetime of our systems undergoing PTDEs. While acknowledging the presence of other potential sources of kicks, such as two-body interactions with surrounding stars, for an upper limit estimation, we only consider the effects of kicks from PTDEs in this paper. These kicks result from the asymmetry in mass loss near the inner and outer Lagrange points L1 and L2 (\cite{manukian2013}; see also Section \ref{sec:ptde}).

For the $i$'th PTDE, the kick velocity of the remnant is labeled by $v_{{\rm kick},i}$. The differences of the specific orbital energy of the remnant between the $i$'th and the $(i-1)$'s PTDEs is given by \citep{manukian2013}
\begin{equation}\label{eq:energy_i}
	\epsilon_{i}-\epsilon_{i-1} = \frac{v_{{\rm{kick}},i}^2}{2}.
\end{equation}
Here, $v_{{\rm{kick}},i}=(0.0745+0.0571\beta_i^{4.539})v_{{\rm esc},i}$; $v_{{\rm esc},i}$ and $\beta_i$ are the escape velocity and the penetration factor of the remnant after the i'th PTDE, respectively. 
If we assume that the final PTDE is marked as $i=N$, then we have 
\begin{equation}\label{eq:energy_n}
	\epsilon_{N}-\epsilon_{N-1} = \frac{v_{{\rm kick},N}^2}{2}.
\end{equation}
According to Equations \ref{eq:energy_i} and \ref{eq:energy_n}, the final specific energy of the remnant is given by 
\begin{equation}
	\label{eq:N}
	\epsilon_{N}-\epsilon_{0} = \sum_{1\le i\le N}\frac{v_{{\rm kick},i}^2}{2}.
\end{equation}
Here, the initial specific orbital energy of the remnant (i.e. the parent star) is $\epsilon_0$. 
If the PTDEs are ended by an ejection, then there is $\epsilon_{N}\approx0$.  

In reality, after each pericenter passage, $\beta_i$ and $v_{{\rm esc},i}$ undergo a few changes due to the remnant's expansion driven by both additional heat and rapid rotation \citep{ryu2020}. However, considering that the orbital period of the remnant around the IMBH is significantly shorter than the photon diffusion timescale, the remnant's structure remains relatively stable \citep{ryu2020}. Thus, as a rough approximation, we assume that $v_{{\rm esc},i} \approx v_{{\rm esc},0}$, where $v_{{\rm esc},0}$ is the initial escape velocity of the parent star.
Furthermore, given that $v_{{\rm kick},i}$ increases with $\beta_i$ and $\beta_{\rm p} \le \beta_i \le \beta_{\rm d}$, we choose an upper limit scenario by assuming $\beta_i=\beta_{\rm p}=0.6$ to estimate the maximum number of PTDEs.

Finally, based on Equation \ref{eq:N}, the maximum number of PTDEs is given by
\begin{equation}
N\sim -\frac{2\epsilon_0}{v_{{\rm kick},0}^2}\approx 363.
\end{equation}
Here, we adopt the following parameters: The initial specific energy of the star around the IMBH is $\epsilon_0=-Gm_2/(2a_\inner)$, where $m_2=10^4\ M_\odot$. Given the eruption period in ASASSN-14ko is $P=115\days$, the semi-major axis of the inner binary is $a_\inner\approx9.97\au$. The initial escape velocity of the star is $v_{{\rm esc},0}= \sqrt{2Gm_\ast/R_\ast}\approx \sqrt{2GM_\odot/R_\odot}$. Therefore, for a period of $P=115\days$, the system can only persist for a maximum timescale of about $114\yr$. 

The actual lifetime of the triple system could be even shorter than what we have estimated. The tidal radius for FTDEs may be slightly enlarged by some factors, such as an increasing mass ratio $m_2/m_1$ or an enlarged remnant radius $R_\ast$. The increasing mass ratio is a result of the mass loss after a PTDE, while the enlarged remnant radius $R_\ast$ is influenced by extra tidal heat and rapid rotation \citep{ryu2020}. Consequently, the remnant could experience a FTDE before the final ejection.

\subsection{The simulation}
We employ N-body code $\textsc{REBOUND}$ \citep{rein2012} equipped with \textsc{IAS15} integrator \citep{rein2015} to simulate our models. The post-Newtonian (PN) terms up to 2.5PN are added in our simulations. Specifically, we account for the periastron shift (1PN and 2PN) and the quadrupole gravitational radiation (2.5PN) based on \cite{kupi2006}, and also the 1PN cross terms as described in \cite{will_14a}, which is resulted from the couplings of potentials between a binary and the third body. The complete form of our accelerations are shown in Appendix \ref{sec:appendix}.

We will terminate our integrations when any of the following conditions are met: (1) a star experiences a partial or full tidal disruption by either black hole; (2) a star becomes gravitationally captured by either black hole; (3) a star is ejected from the SMBH-IMBH binary, when the star has a positive orbital energy relative to the SMBH and also has a relative distance from the center of mass of the SMBH-IMBH binary greater than $50a_\out$. The maximum integration timescale is set to $10^6$ years.

In this study, we employ high-accuracy N-body simulations to investigate various dynamical processes in the triple system. These mechanisms include the LK mechanism, the periastron shift (1PN and 2PN), the quadrupole gravitational radiation (2.5PN), the three-body interaction, and the 1PN cross terms.

\begin{figure}
	\includegraphics[width=1\linewidth]{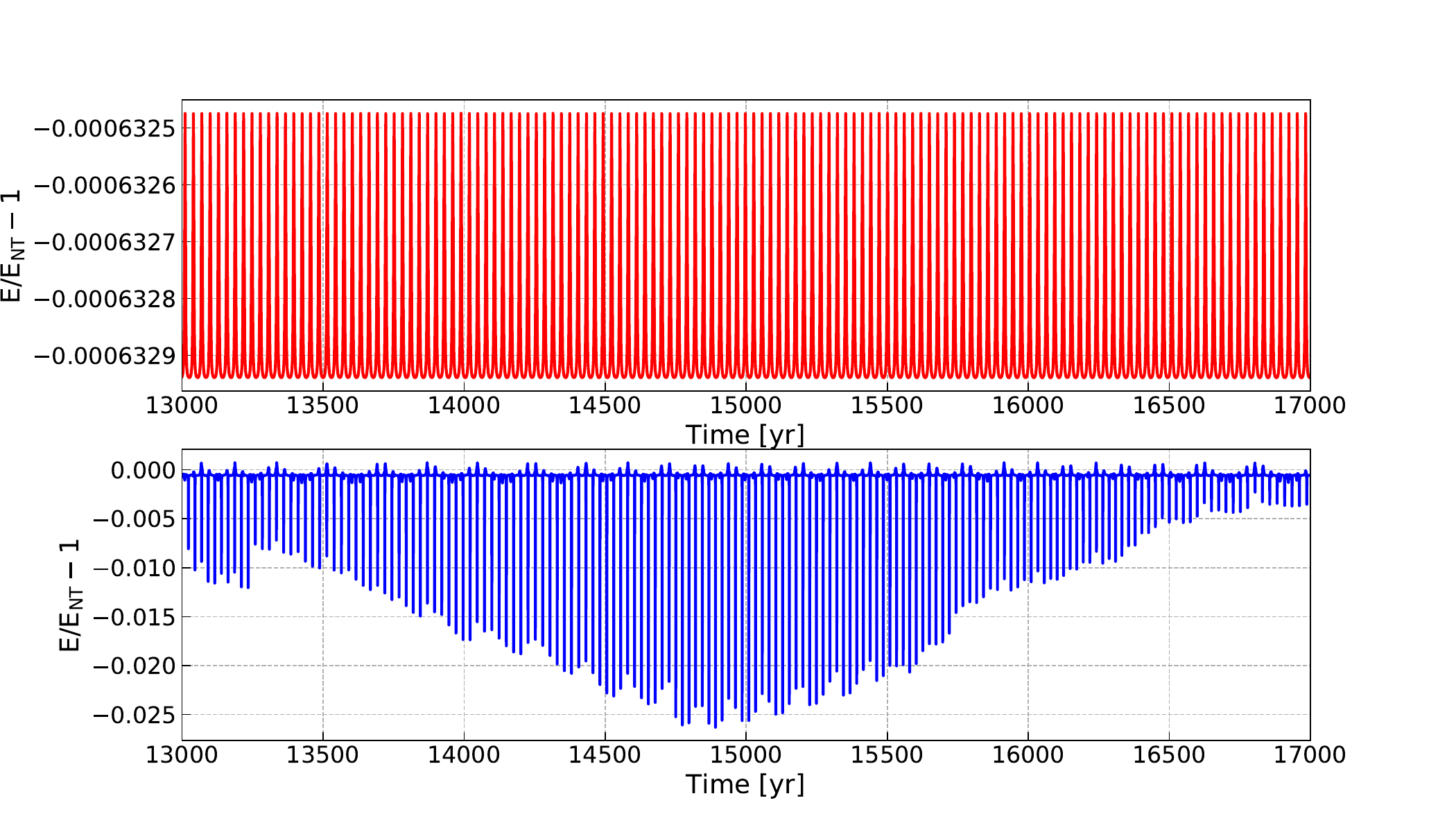}
	\caption{An illustration of the relative ratio of the 1PN orbital energy to the Newtonian orbital energy for one of our triple systems. $E_{\rm NT}$ is the Newtonian orbital energy, representing the sum of the kinetic energy and Newtonian potential energy. In the \textit{top panel}, all 1PN terms, including those related to the periastron shift and cross terms, are incorporated. The 1PN orbital energy is calculated using Eq. 3.2a from \citet{will_14a}. In the \textit{bottom panel}, only the 1PN term concerning the periastron shift is considered, and the 1PN orbital energy is computed using Eq. 11 from \citet{will_14b}. The primary parameters are $m_3=10^7\ M_\odot$, $m_2=10^4\ M_\odot$, $a_\out=0.01\pc$, $e_\out=0.5$, $a_\inner=45.57\au$ and $e_\inner=0.177$.}
	\label{fig:energy_1pn}
\end{figure}

It's important to highlight that the inclusion of certain PN terms, such as the periastron shift of 1PN, does not conserve energy, as demonstrated in \citet{rodriguez2018}. However, \citet{will_14b} has shown that if the cross terms of the 1PN are also incorporated into the dynamical equations, the orbital energy is conserved in the triple systems, as depicted in Fig. \ref{fig:energy_1pn}. 

In Fig. \ref{fig:energy_1pn}, we present the relative ratio of the 1PN orbital energy to the Newtonian orbital energy for one of our triple systems with $m_3=10^7\ M_\odot$, $m_2=10^4\ M_\odot$, $a_\out=0.01\pc$, $e_\out=0.5$, $a_\inner=45.57\au$, and $e_\inner=0.177$. In the top panel, all 1PN terms, including those related to the periastron shift and cross terms, are incorporated. We observe that the 1PN orbital energy is evidently conserved to Newtonian order, as initially proven by \cite{will_14b}. However, in the bottom panel, we notice that the 1PN orbital energy increases by nearly $2.5\%$ when only the 1PN term concerning the periastron shift is considered, as demonstrated by \citet{rodriguez2018}. 
Although a detailed analysis of the effects of the non-conservative energy in triple systems is not within the scope of this article, our findings suggest that, for accurately simulating the orbit of a star around a SMBHB, it may be necessary to account for the full PN corrections. At the very least, incorporating the 1PN cross terms is essential. In our upcoming paper (Wu et al., in preparation), we will delve into the effects of the 1PN cross terms in triple systems.

\section{Results} \label{sec:results}

\begin{table*}
	\caption{Results of simulations: partial tidal disruption events by the IMBH (PTDE1), in which the subscripts "$\rm b$" and "$\rm e$" mean bound and ejection at the time of disruption, respectively; partial tidal disruption events by the SMBH (PTDE2); full tidal disruption events by the IMBH (FTDE1); full tidal disruption events by the SMBH (FTDE2); stars captured by the IMBH (Bound1); stars captured by the SMBH (Bound2); and stars ejected by the SMBHB (Ejection).}
	\centering
	\begin{tabular}{c c c c c c c c c}
	\hline\hline
	Model & PTDE1$_{\rm b}$& PTDE1$_{\rm e}$ & PTDE2 & FTDE1  & FTDE2 & Bound1 & Bound2 & Ejection
	\\[0.5ex]
	\hline
	M01  & 0.1041  & 0.0123  & $-$        & 0.0239  & $-$        & 0.1866& 0.6700 & 0.0031\\
	M02 & 0.2398 & 0.1114    & 0.0141 & 0.1058  & $-$        &  0.1241 & 0.3117  &0.0931 \\
	M03 & 0.1253  & 0.0004 & $-$         & 0.0571  & $-$       & 0.2532 & 0.2781 & 0.2858\\
	M04 & 0.0825 & 0.0039 & 0.0547 & 0.0254 & 0.0046 & 0.2143 & 0.3554 & 0.2591\\
	M05 & 0.1044 & 0.0009 & 0.0815 & 0.0321 & 0.0046 & 0.1841 & 0.2967 &0.2958\\
	M06 & 0.1164 & 0.0007 & 0.0539 & 0.0226 & 0.0166 & 0.2069 & 0.0013 & 0.5815\\
	\hline 
	\end{tabular}
	\label{tab:results}
\end{table*}

Table \ref{tab:results} presents the results of our simulations, including the ratios of PTDEs, FTDEs, bound cases, and ejection cases. For PTDEs, we differentiate between those caused by the IMBH and those caused by the SMBH. We determine if the remnants of stars after the first partial disruption can be ejected from the IMBH based on their orbital energies relative to the IMBH.

Our findings indicate that the majority of PTDEs are generated by the IMBH. After the initial disruption, a significant portion of these PTDEs are subsequently captured by the IMBH, suggesting the possibility of a second flare. In addition, we observe a small number of PTDEs that are disrupted by the SMBH, and all of these PTDEs are captured by the SMBH. Furthermore, alongside the PTDEs, FTDEs are present in all models.
Moreover, it is worth noting that when $m_2<10^5\rm\ M_\odot$ (as seen in models from M01 to M05), the majority of stars are bound to either the IMBH or the SMBH. This dependency is determined by the mass of the IMBH and the semi-major axis of the outer binary. However, in the case of a massive IMBH (e.g., in model M06 where $m_2=10^5\rm\ M_\odot$), the majority of stars are instead ejected away from the SMBHB.

\subsection{The captured stars}
\begin{figure}
	\includegraphics[width=1\linewidth]{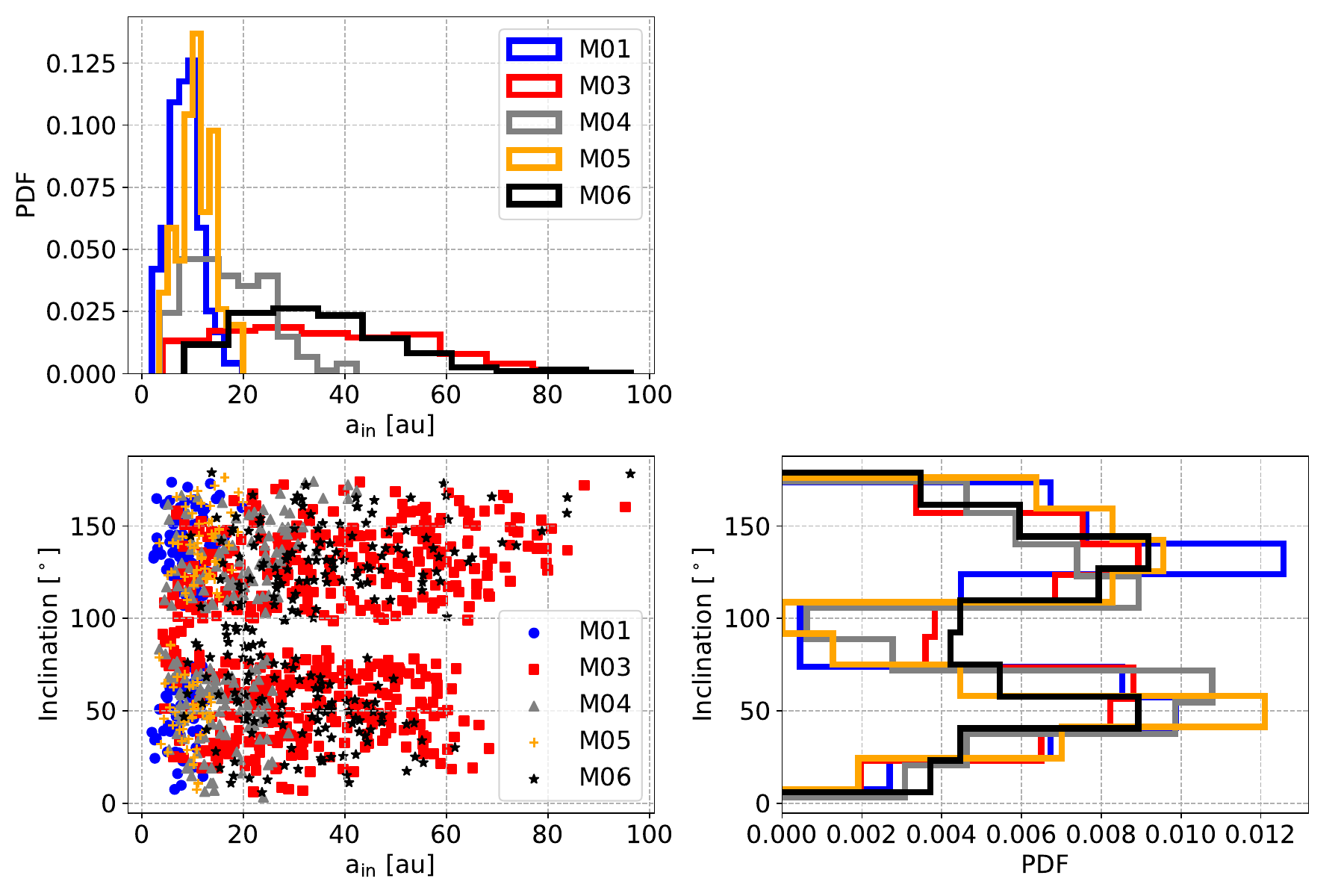}
	\caption{The final distributions of the captured stars by the IMBH after $1\myr$ in the semi-major axis - inclination space. The blue, red, gray, orange and black points and lines are for the models M01, M03, M04, M05 and M06.
	\textit{Bottom left panel}: distributions of bound stars around the IMBH in the space of the semi-major axis and inclination.
	\textit{Top left panel}: probability distribution functions (PDFs) of the semi-major axis of bound stars. \textit{Bottom right panel}: PDFs of the inclination angle between the orbital planes of the inner and outer binaries.}
	\label{fig:bound1}
\end{figure}

Figure \ref{fig:bound1} presents the distributions of bound stars around the IMBH in our models. The bottom left panel displays the distributions of bound stars in the space of the semi-major axis and inclination between the orbital planes of the inner and outer binaries. The top left panel shows the probability distribution functions (PDFs) of bound stars with the semi-major axis, while the bottom right panel depicts the PDFs of bound stars with the inclination angle.

In all models, we observe a torus structure around the IMBH, which is attributed to the Lidov-Kozai mechanism \citep{lidov1962,kozai1962,naoz2016,fragione2018}. This torus structure was initially proposed by \cite{li2015} and subsequently confirmed through simulations by \cite{fragione2018}.
By comparing the bound stars in models M03, M04, and M05, as depicted in the top left panel of Figure \ref{fig:bound1}, it becomes evident that the eccentricity of the outer binary can narrow the PDF of bound stars with the semi-major axis. Furthermore, a comparison of the bound stars in models M01, M04, and M06 reveals that as the IMBH becomes more massive, the PDF with the semi-major axis becomes much wider.

\subsection{Partial tidal disruption events}

The period of a binary system can be calculated by
\begin{equation}
	P=2\pi \sqrt{\frac{a^3}{GM}}=\left(\frac{a}{\rm{1\ au}}\right)^{3/2}\left(\frac{M}{1\rm\ M_\odot}\right)^{-1/2}\yr,
\end{equation}
where $G$ is the Newtonian constant, and $a$ and $M$ are the semi-major axis and total mass of the binary system, respectively. If we assume the binary system is the inner binary of our models, and has the period of $P=115\days$, which is the period of  the quasi-periodic eruptions in ASASSN-14ko \citep{payne2021}, and $M_\inner=10^3, 10^4$, and $10^5\rm\ M_\odot$, which is the mass of the IMBH, then the semi-major axis of the inner binary is $a_\inner\approx 4.63, 9.97$, and $21.48\au$, respectively.

\begin{figure}
	\centering
	\includegraphics[width=\linewidth]{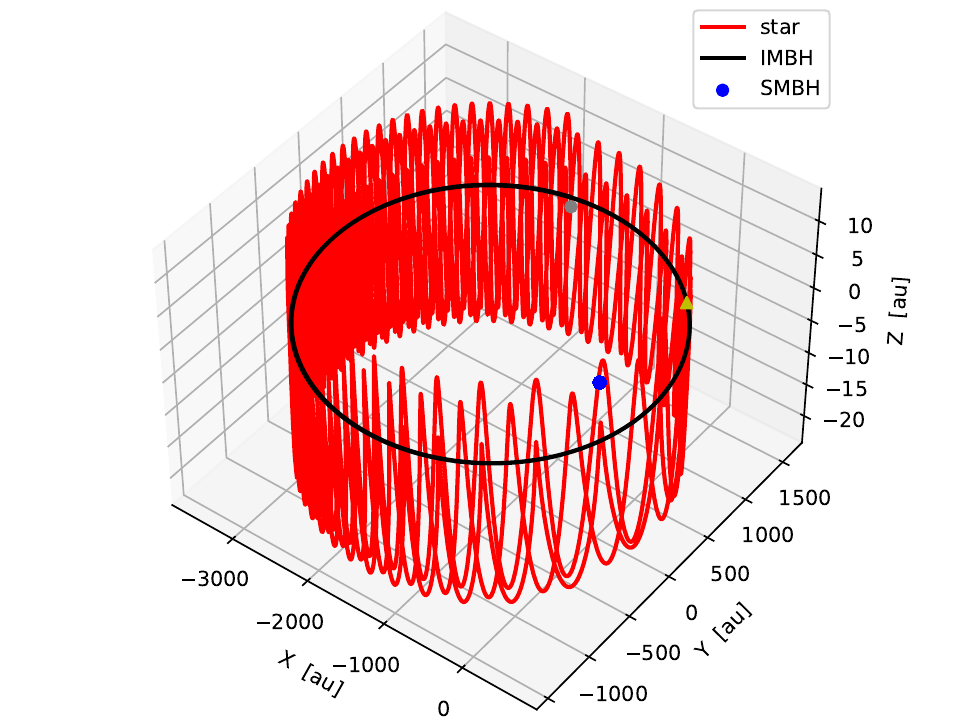}
	\includegraphics[width=0.9\linewidth]{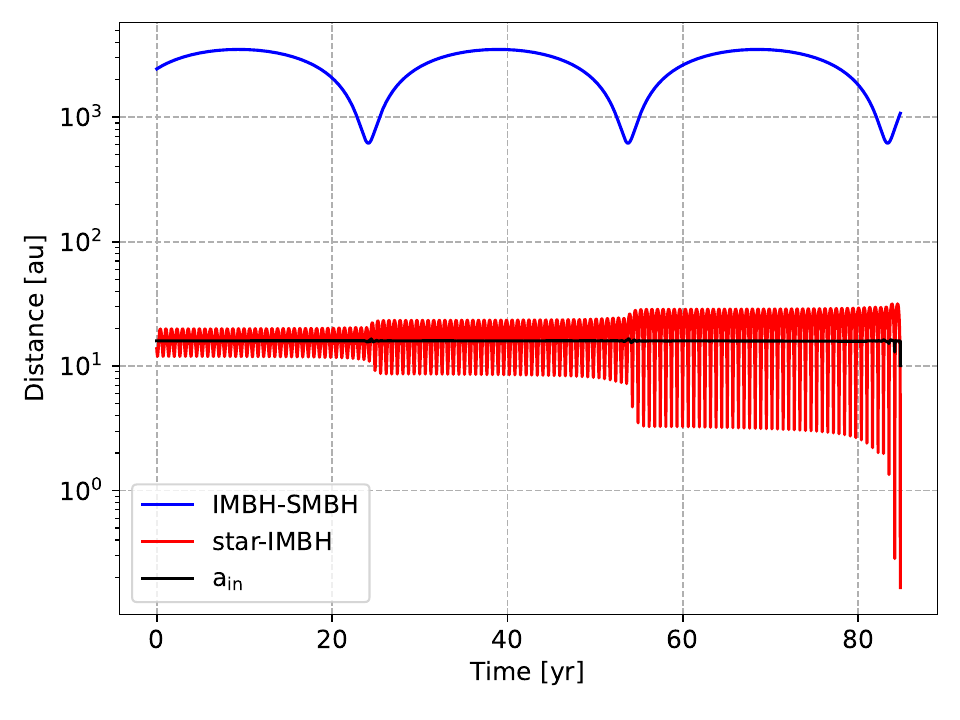}
	\caption{\textit{Top panel}: An illustration of the 3D orbit of a star that undergoes a partial disruption by the IMBH. The origin of the frame of the reference is the center of mass of the system. The red, black lines represent the trajectories of the star, IMBH, respectively. The blue points represent the orbit of the SMBH. The gray point marks the initial position of the star, while the yellow triangle indicates the location of the disruption by the IMBH. 
\textit{Bottom panel}: The evolutions about the distance of the star around the IMBH (red line), the distance of the IMBH around the SMBH (blue line), and the semi-major axis of the star-IMBH binary (black line), as a function of time, corresponding to the example shown in the top panel. The main parameters are as follows: $m_3=10^7\rm\ M_\odot$, $m_2=10^4\rm\ M_\odot$, $a_\out=0.01\pc$, $e_\out=0.7$, $a_\inner=15.946\au$, and $e_\inner=0.248$. The final orbital period of the star at the time of the PTDE is $117.24\days$.}
	\label{fig:example}
\end{figure}

Figure \ref{fig:example} illustrates an example of a PTDE induced by the IMBH. We selected specific parameters for this scenario: $m_3=10^7\rm\ M_\odot$, $m_2=10^4\rm\ M_\odot$, $a_\out=0.01\ \rm pc$, $e_\out=0.7$, $a_\inner=15.946\ \rm au$, and $e_\inner=0.248$. 
In the top panel, the 3D orbits of the star, IMBH and SMBH are depicted, where the star's orbit around the IMBH is approximately perpendicular to the orbit of the outer binary. 
The bottom panel presents the evolution of some key parameters of the system, including the distance of the star around the IMBH, the semi-major axis of the star-IMBH binary (i.e. $a_\inner$), and the distance of the IMBH around the SMBH. The semi-major axis of the star remains almost constant, in accordance with the descriptions of the LK mechanism. On the other hand, slight variations about $a_\inner$ occur when the IMBH crosses the pericenter around the SMBH, which results from the strong interactions among the triple system. At the moment of disruption, the star reaches a final semi-major axis of $10.1\ \rm au$, corresponding to an orbital period of $117.24\days$.

\begin{figure}
	\includegraphics[width=1\linewidth]{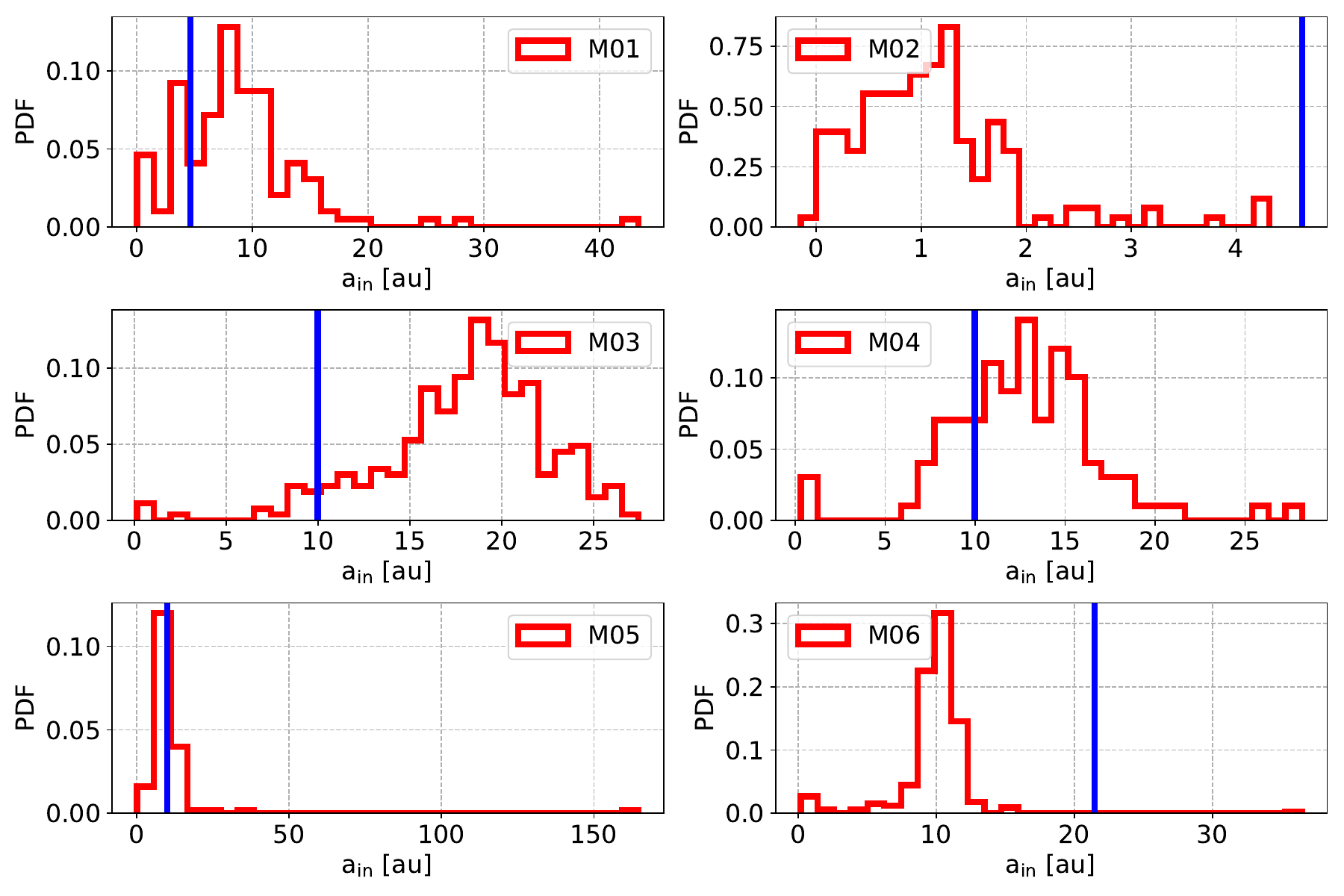}
	\centering
	\caption{The probability distributions of PTDEs caused by the IMBH are shown, illustrating the distribution of the semi-major axis of the inner binary at the time of disruption. PTDEs with a positive orbital energy have been excluded from the analysis. In each panel, the vertical blue line represents the orbit of a star with an orbital period of $115\days$, which corresponds to the period of QPEs observed in ASASSN-14ko \citep{payne2021}.}
	\label{fig:ptde1}
\end{figure}

Figure \ref{fig:ptde1} presents the PDFs of PTDEs caused by the IMBH, specifically showing the distribution of the semi-major axis of the inner binary at the time of disruption. In each panel, the vertical blue line represents a stellar orbital period of $115\days$, which corresponds to the period of the eruptions observed in ASASSN-14ko \citep{payne2021}. We observe that the PTDEs in models M01, M04, and M05 align with the period observed in ASASSN-14ko. However, when we introduce variations to certain system parameters, such as the mass of the IMBH (e.g., model M06), the semi-major axis of the outer binary (e.g., model M02), or the eccentricity of the outer binary (e.g., model M03), the resulting PDFs no longer match the observed period.

Based on our findings, we propose that there might be a SMBHB at the center of the host galaxy of ASASSN-14ko, possessing the following characteristics: the primary black hole has a mass of approximately $10^7\rm\ M_\odot$, the secondary black hole has a mass less than $\sim10^5\rm\ M_\odot$, the eccentricity of the binary is greater than or equal to $\sim0.5$, and the semi-major axis of the binary is $\sim0.01\pc$.

\subsection{The period derivative}

In this section, we explore the possibility that the period derivative ($\dot{P}$) of the eruptions in ASASSN-14ko, observed to be $\dot{P}=-0.0026\pm 0.0006$ \citep{payne2023}, results from the variation in the arrival time of light from each eruption position to the observer on Earth. Due to the orbital motion of the IMBH around the SMBH, the arrival time of light from each eruption differs, leading to a non-constant period of eruptions.

\begin{figure}
	\centering
	\includegraphics[width=0.9\linewidth]{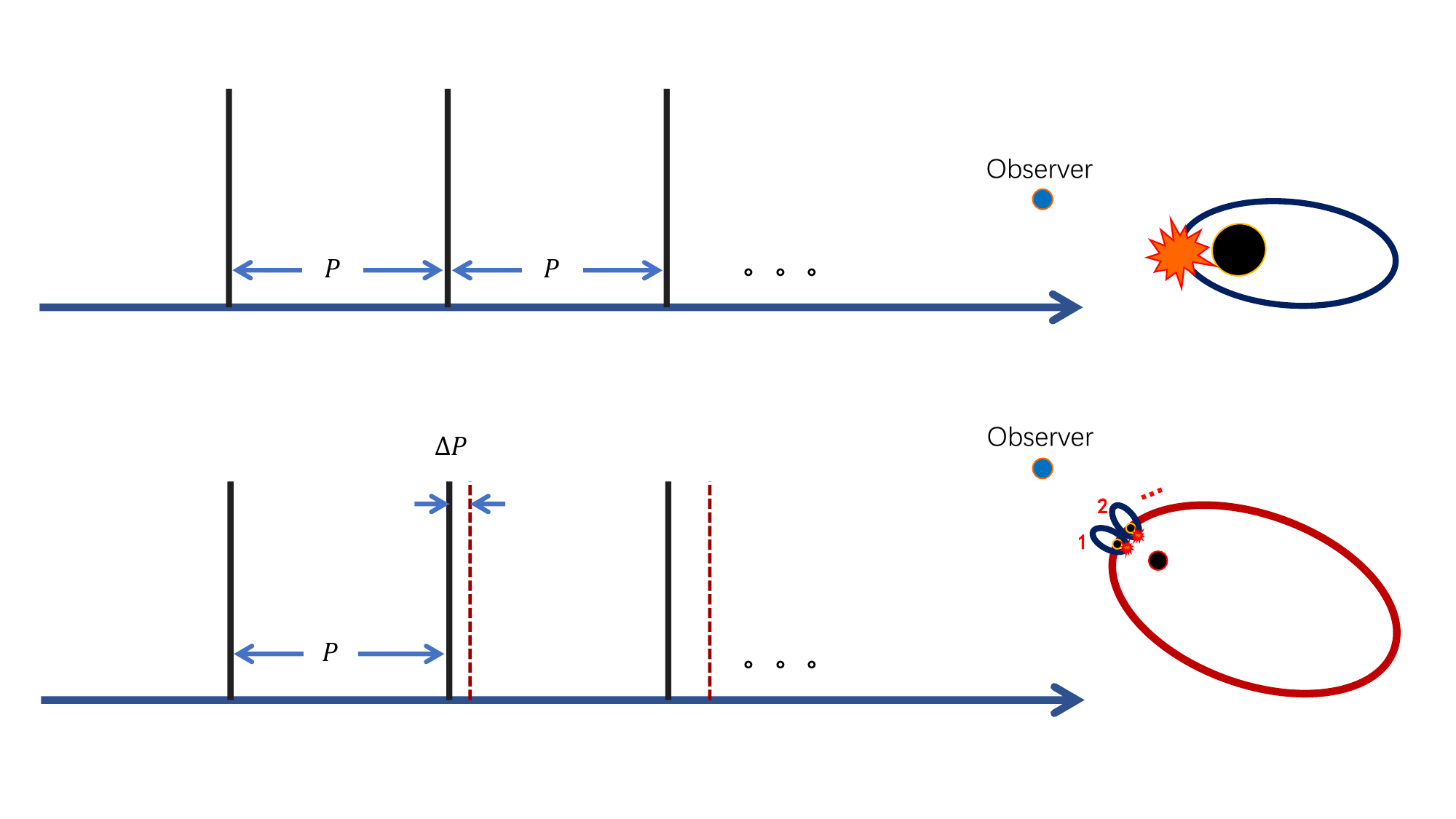}
	\caption{An illustration depicting the change in the period ($\Delta P$) of periodic eruptions. The blue point marks the observer on Earth. \textit{Top panel} illustrates the eruptions with a constant period in a star-SMBH binary system, assuming the absence of some dissipative effects such as gravitational-wave radiation and tidal dissipation. \textit{Bottom panel} depicts the eruptions with a varying period in our triple system. Here, the varying period results from the orbital motion between the two massive black holes. In the \textit{bottom panel}, we assume that the IMBH is moving away from our vantage point, causing the period to increase over time.}
	\label{fig:fig1}
\end{figure}

Fig. \ref{fig:fig1} illustrates a schematic representation of the relationship between the IMBH's motion and the variations in eruption period. For comparison, the top panel of Fig. \ref{fig:fig1} shows the variations in the eruption period in the star-SMBH binary system. It is evident that, in the absence of dissipative factors like gravitational-wave radiation and tidal dissipation, the period between successive eruptions ($P$) remains nearly constant due to the almost constant arrival time of light from each eruption. While the periastron shift of the remnant around the SMBH could induce some variations in the period by changing the arrival time of light, we observe that the shift-related effect is minor.

However, the bottom panel of Fig. \ref{fig:fig1} reveals that the period is no longer constant but instead varies with the IMBH's motion around the SMBH. When the IMBH moves away from the observer, the distance between the eruption position and the observer increases, leading to a lengthening of the period over time ($\dot{P} > 0$). Conversely, as the IMBH orbits towards the observer, the period decreases over time ($\dot{P}< 0$). As a result, the sign of $\dot{P}$ undergoes an approximately periodic change every $P/2$.

\begin{figure}
	\centering
	\includegraphics[width=0.9\linewidth]{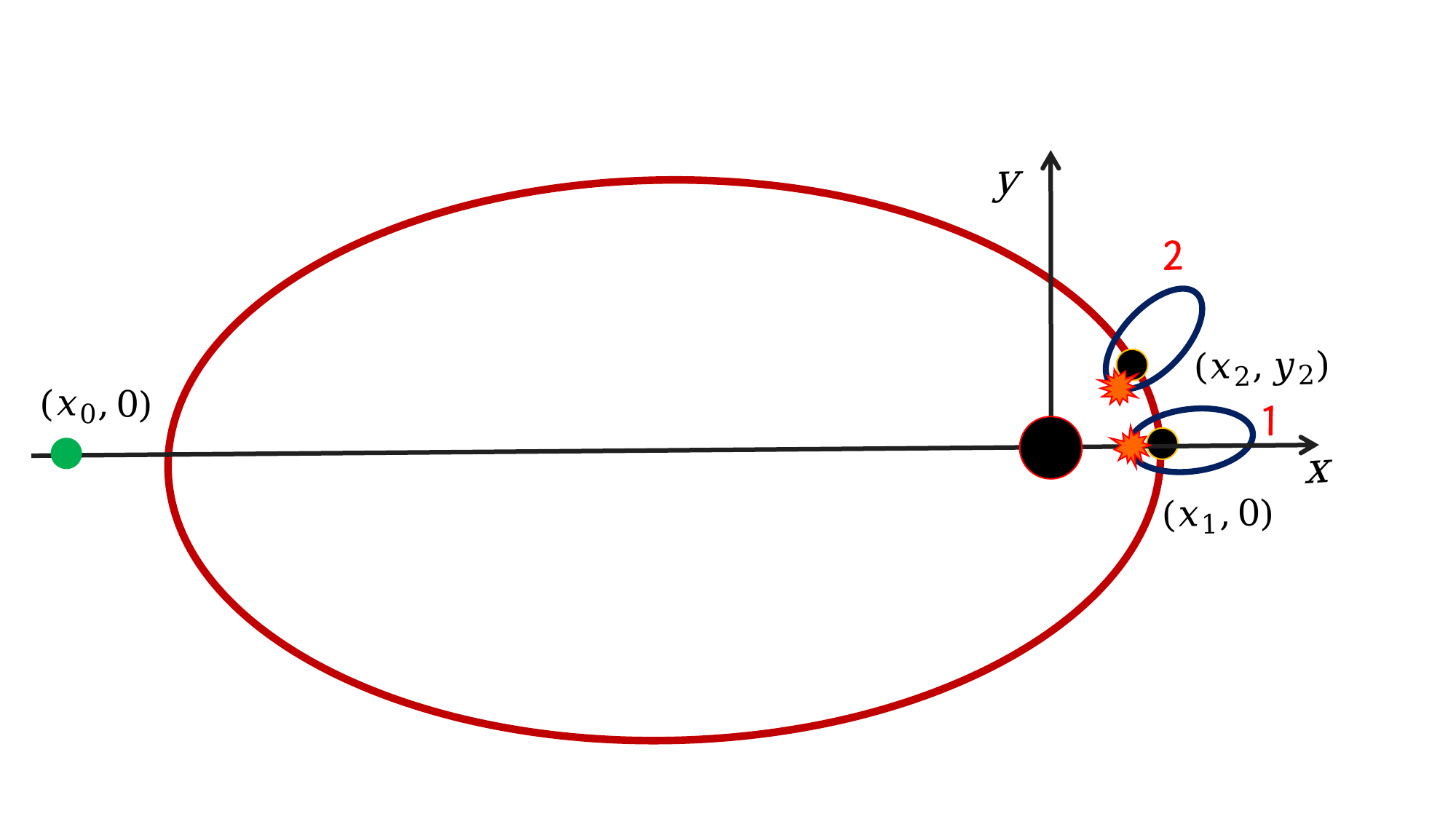}
	\caption{An illustration of the model used in this section, although not to scale, is presented here. In this representation, the SMBH is situated at the origin of the frame, the x-axis aligns with the semi-major axis, and the y-axis is parallel to the semi-minor axis. Two successive eruptions are denoted as 1 and 2. The coordinates of the observer, the eruption 1, and the eruption 2 are represented as $(x_0,0)$, $(x_1, 0)$, and $(x_2, y_2)$, respectively. For the sake of simplification, we assume that the coordinates of the star and the IMBH are identical, owing to their close proximity relative to the location of the observer.}
	\label{fig:fig2}
\end{figure}

As illustrated in Fig. \ref{fig:fig2}, we define a coordinate system with the origin fixed at the location of the SMBH. In this frame, the x-axis aligns with the semi-major axis of the SMBH-IMBH binary, while the y-axis aligns with the semi-minor axis of the SMBH-IMBH binary. The positions of the observer and two successive eruptions are denoted as $(x_0, 0)$, $(x_1, y_1)$, and $(x_2, y_2)$, respectively. 

For the sake of simplification, we assume that the eruptions initially occur at the perihelion of the IMBH around the SMBH, with the observer situated along the x-axis to the right of the SMBH. Consequently, for the initial eruption (i.e. the eruption 1 in Fig. \ref{fig:fig2}), we have $y_1=0$. Furthermore, it's essential to clarify that we consider the coordinates of the star and IMBH to be identical due to their close proximity during each eruption, and the periastron shift is not considered in this section as its effects are minor.

In this model, the period derivative is estimated as follows:
\begin{equation}
\frac{d P}{d t} \approx \frac{\Delta P}{P} = \frac{\Delta d}{c P}.
\end{equation}
Here, $\Delta P$ represents the change in the period of periodic eruptions, and $\Delta d$ corresponds to the change in the distance between an eruption and the observer, which results from the IMBH's motion around the SMBH.

The orbit of the IMBH can be described by the equation \citep{landau1976}:
\begin{equation}
	r=a_\text{out}(1-e_\text{out}\cos{E}).
\end{equation}
Here, $E$ represents the eccentric anomaly, which ranges from $0$ to $2\pi$.

The time elapsed since the passage of the perihelion can be given by:
\begin{equation}\label{eq:t}
	t=\sqrt{\frac{\mu a_\text{out}^3}{Gm_2m_3}}(E-e_\out\sin{E}),
\end{equation}
In this equation, $\mu=m_2m_3/(m_2+m_3)$.

The coordinates of the IMBH can then be calculated using the following expressions:
\begin{align}\label{eq:imbh_coordinate}
	& x = a_\text{out}(\cos{E}-e_\text{out}),\\ \notag
	& y = a_\text{out}\sqrt{1-e_\text{out}^2}\sin{E}. 
\end{align}
Here, $E$ is calculated by Equation \ref{eq:t}.

Finally, the distance between the eruption 2 and the observer is determined by:
\begin{equation}
	d_1 = \sqrt{(x_2-x_0)^2+y_2^2},
\end{equation}
where $x_2$ and $y_2$ are the coordinate of the IMBH, which can be calculated by Equation \ref{eq:imbh_coordinate}. 
Therefore, we have $\Delta d=d_1-d_0$, where $d_0=188\ \text{Mpc}$ \citep{payne2022a} represents the distance between the initial eruption and the observer.

As we've assumed that the initial eruption takes place at perihelion, for the first two eruptions in Fig. \ref{fig:fig2} (i.e., eruptions 1 and 2), we can set $t=P$, where $P=115\days$ \citep{payne2021,payne2022a}. Thus, for the eruption 2, according to Equations \ref{eq:t}, we can approximate $E\approx 0.219$. If we further assume that the SMBH-IMBH binary is from model M05 (as detailed in Table \ref{tab:initial_conditions}), then the following parameters apply: $x_1=a_\out(1-e_\out)=3\times10^{-3}$ pc, $x_0=-d_0+x_1$, $x_2=a_\out(\cos{E}-e_\out)\approx 2.76\times10^{-3}$ pc, and $y_2=a_\out\sqrt{1-e_\out^2}\sin{E}\approx 1.55\times10^{-3}$ pc.

Finally, in our configuration (see Fig. \ref{fig:fig2}), we have the value of 
\begin{equation}
	\frac{dP}{dt} \approx -0.0025, 
\end{equation}
which matches the value of $\dot{P}=-0.0026\pm 0.0006$ in ASASSN-14ko \citep{payne2023} very well.

\begin{figure}
	\centering
	\includegraphics[width=1\linewidth]{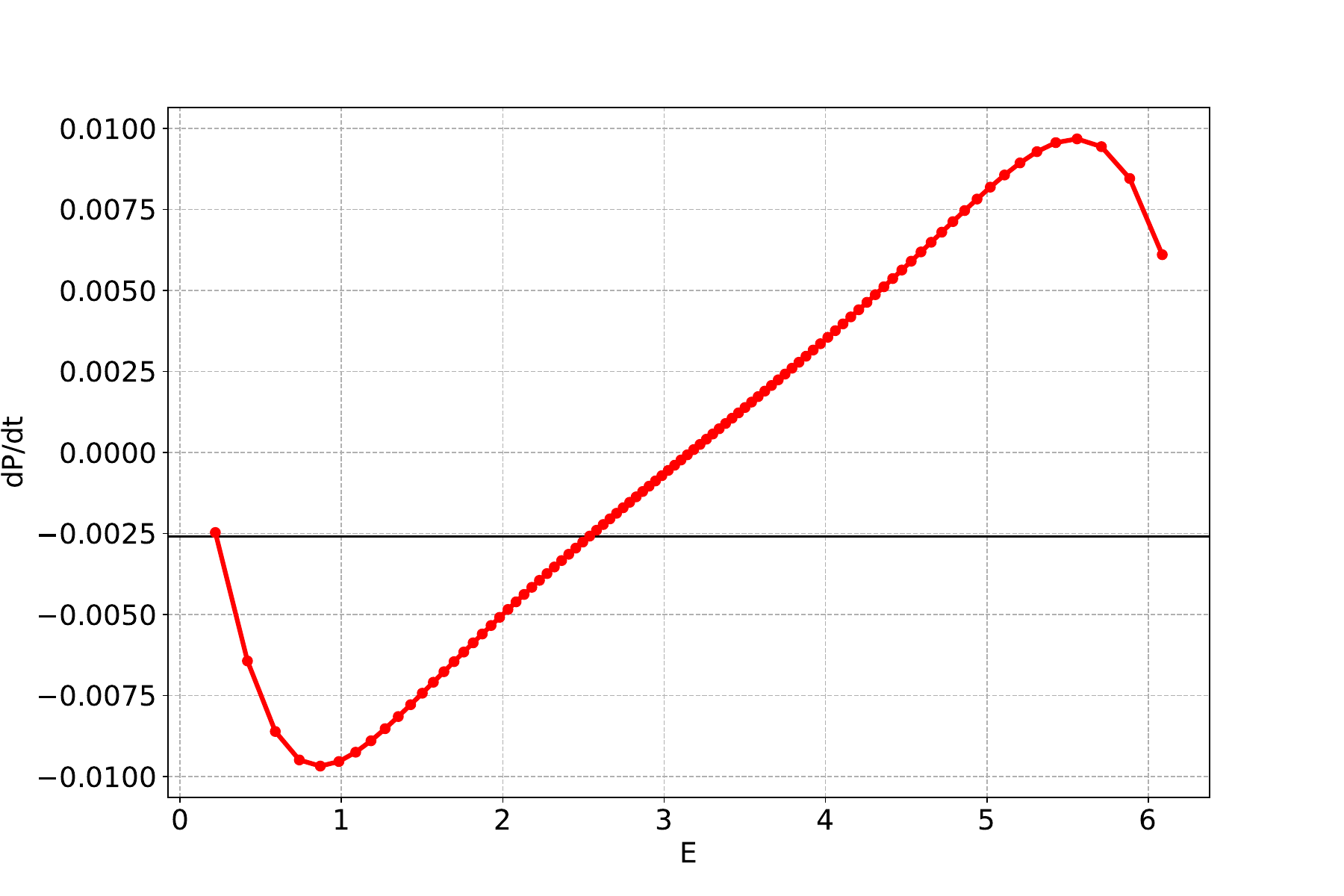}
	\caption{The evolution of the period derivative ($\dot{P}$) along the orbit of the IMBH around the SMBH is shown as a function of the eccentric anomaly ($E$) of the orbit of the IMBH. The black horizontal line represents the observational value of $\dot{P}=-0.0026\pm 0.0006$ in ASASSN-14ko \citep{payne2023}. Here, we have assumed that the first eruption happens when the IMBH is at the perihelion around the SMBH, as illustrated in Fig. \ref{fig:fig2}.}
	\label{fig:fig3}
\end{figure}

Figure \ref{fig:fig3} shows the evolution of the period derivative of the eruptions with the motion of the IMBH around the SMBH. In this plot, the black horizontal line represents the observed value of $\dot{P}=-0.0026\pm 0.0006$ in ASASSN-14ko \citep{payne2023}. Notably, we find that for certain specific eccentric anomalies, our model is capable of explaining the observed derivative. Furthermore, we also note that $\dot{P}$ varies with the IMBH's position. Specifically, when the IMBH approaches us, $\dot{P}$ becomes negative, while $\dot{P}$ becomes positive when the IMBH moves away from us. 

It's important to note that our model is a simplified representation. In addition to the discussed position of the IMBH relative to the SMBH, the actual model is influenced by various other factors. These include the orientation of the orbital plane of the SMBH-IMBH binary relative to the line of sight of the observer, the orientation of the orbital plane of the star-IMBH binary relative to the orbital plane of the SMBH-IMBH binary, the periastron shift of the star around the IMBH, and the kick velocity imparted to the remnant of the star after each PTDE. Therefore, a comprehensive model should account for all these factors.

\section{Conclusion and Discussions} \label{sec:conclusions_and_discussions}
In galactic nuclei, extensive researches have been conducted on the tidal disruptions of stars around SMBHBs \cite[e.g.][]{chen2009,chen2011,wegg2011,li2015,fragione2018}. However, there have been limited investigations into the phenomenon of stars being partially disrupted by SMBHBs. The presence of the Lidov-Kozai mechanism \citep{lidov1962,kozai1962,naoz2016} allows stars in SMBHB systems to occupy a region where partial tidal disruptions by the SMBHBs are possible \citep{guillochon2013,melchor2023}.

This study focuses on exploring partial tidal disruptions of stars orbiting an IMBH in the presence of an outer SMBH. We aim to compare our results with the observed period of the periodic eruptions in ASASSN-14ko. To achieve this, we consider a fixed SMBH mass of $10^7\rm\ M_\odot$, which is close to the mass of the SMBH in ASASSN-14ko \citep{payne2021}, and systematically vary the mass of the IMBH, as well as the semi-major axis and eccentricity of the SMBH-IMBH binary.

Indeed, IMBHs are believed to be common in galactic centers. Numerous studies have suggested the presence of an IMBH in our own galactic center \cite[e.g.,][]{hansen2003,yu2003,gualandris2009,chen2013,naoz2020}. Several scenarios have been proposed to explain the formation of IMBHs in galactic centers. For example, research by \cite{rose2022} indicates that stellar black holes could grow in mass through repeated collisions with stars in galactic centers, resulting in IMBHs with masses reaching up to $10^4\rm\ M_\odot$. Additionally, IMBHs can be formed in globular clusters and later migrate into galactic centers through dynamical processes \cite[e.g.,][]{gnedin2014,arca-sedda2018,fragione2022b}. Moreover, SMBH-IMBH binaries could form through minor mergers between massive and low-mass galaxies, provided that low-mass galaxies harbor IMBHs at their centers \cite[e.g.,][]{moran2014,greene2020}.

In summary, our study yields two key findings: 1) Our model reveals that a specific configuration of a SMBH and an IMBH in a binary system can generate PTDEs with a periodicity of approximately $115\days$, akin to the periodic eruptions observed in ASASSN-14ko. This configuration requires a primary SMBH mass of $10^7 M_\odot$, a secondary IMBH mass below $\sim 10^5 M_\odot$, a binary eccentricity equal to or greater than $\sim0.5$, and a binary semi-major axis exceeding $0.001\pc$. 2) Our model not only accounts for the observed period derivative of the periodic eruptions in ASASSN-14ko ($\dot{P}=-0.0026\pm 0.0006$) \citep{payne2023}, but also predicts that the period derivative evolves with the motion of the IMBH, transitioning from decay ($\dot{P}<0$) to growth ($\dot{P}>0$) approximately every $P/2$.

Distinguishing between PTDEs originating from the Hills mechanism and those arising from the Lidov-Kozai mechanism using hydrodynamic simulations is extremely challenging. To date, no hydrodynamic simulations have encompassed the entire evolutionary stages of PTDEs in star-SMBH binaries, let alone PTDEs in SMBH-IMBH binaries. These systems involve numerous active factors, such as tidal dissipation \citep{press1977}, post-Newtonian effects \citep{kupi2006,will_14a}, Lidov-Kozai effects \citep{lidov1962,kozai1962,naoz2016}, and chaotic orbital perturbations \citep{chen2011}. Consequently, comprehensive hydrodynamic simulations capable of capturing all these complexities are currently unavailable.

Indeed, \cite{mockler2023} suggest that by comparing the mass derived from the light curve of a PTDE to the SMBH mass estimated from the galaxy scaling relation \citep{tremaine2002}, it may be possible to identify the presence of a SMBHB at the galactic center. If the derived black hole mass is smaller than the expected SMBH mass, it could indicate the existence of a secondary black hole in the system, supporting the presence of a SMBHB. This approach provides a potential method of distinguishing between different mechanisms for PTDEs and inferring the presence of binary black holes in galactic nuclei.

It is worth noting that in model M02 (see Table \ref{fig:ptde1}), the peak of the PDF corresponds to a semi-major axis of the inner binary $a_\inner \sim 1\au$, which corresponds to an orbital period of approximately $2.77\rm\ h$. This timescale appears to be consistent with the recurrence time observed in some QPEs \citep{king2022}. A detailed analysis of our model specifically focused on QPEs will be presented in a forthcoming paper. 

In addition to the PTDEs discussed in this paper, SMBHBs can efficiently produce extreme-mass-ratio inspirals (EMRIs) through the capture of stellar-mass compact objects by SMBHs \cite[e.g.,][]{bode2014,naoz2022,mazzolari2022}. According to these studies, the LK mechanism plays a crucial role in the formation of EMRIs. TDEs (both partial and full) and EMRIs share a common formation channel in SMBHBs, as they are brought to critical distances from SMBHs primarily due to the LK mechanism. Therefore, in our system, EMRIs around SMBHBs are expected to exhibit some characteristics similar to those of PTDEs, such as the orbits of compact objects around the IMBH (see Fig. \ref{fig:example} for reference) and the PDFs of the semi-major axis around the IMBH (see Fig. \ref{fig:ptde1} for reference). \citet{mazzolari2022} obtained similar distributions of initial relative inclination and final semi-major axis for EMRIs as observed in our PTDEs, although in their study, the outer SMBH is the less massive secondary. A detailed analysis is beyond the scope of this study, and we will study their effects in our future work.

\section*{Acknowledgements}
We thank the referee for the important and insightful comments which are very helpful in the improvement of this work. 
This work was supported by 
National Natural Science Foundation of China (Grant No. 12393812 and 11973025),
National SKA Program of China (Grant No. 2020SKA0120300) and 
the Strategic Priority Research Program of the Chinese Academy of Sciences (Grant No. XDB0550200). 
\section*{Data Availability}
The data underlying this article will be shared on reasonable request to the corresponding authors (Ye-Fei Yuan and Wenbin Lin).



\bibliographystyle{mnras}
\bibliography{references} 



\appendix
\appendix
\onecolumn
\section{The Post-Newtonian accelerations}\label{sec:appendix}
The acceleration of the body $i$ is calculated as follows,
\begin{equation}
	\bm{a}=\bm{a}_0+\bm{a}_{\rm 1PN} + \bm{a}_{\rm 2PN} + \bm{a}_{\rm 1PNc} + \bm{a}_{\rm 2.5PN},
\end{equation}
where $\pmb{a}_0$ is Newtonian acceleration; $\pmb{a}_{\rm 1PN}$ and $\pmb{a}_{\rm 2PN}$ are 1PN and 2PN accelerations respectively, which are responsible for the pericenter shift \citep{kupi2006}; $\pmb{a}_{\rm 2.5PN}$ is 2.5PN acceleration, which is responsible for the quadrupole gravitational radiation \citep{kupi2006}; $\bm{a}_{\rm 1PNc}$ is the acceleration of the 1PN cross terms, which is resulted from the couplings of potentials between the inner binary and the third body \citep{will_14a}.

The different terms of the acceleration of the body $i$ are calculated as follows,
\begin{equation}
	\pmb{a}_0 = -\sum_{j\neq{i}}\frac{Gm_j\pmb{n}_{ij}}{r_{ij}^2}.
\end{equation}

\begin{equation} \label{eq:2pn}
	\pmb{a}_{\rm 1PN} = \frac{1}{c^2} {\sum_{j\neq{i}}}\frac{G m_{j}\pmb{n}_{ij}}{r_{ij}^2} \bigg[ 4\frac{G m_j}{r_{ij}}+5\frac{G m_i}{r_{ij}}-v_{i}^2+4\pmb{v}_{i}\cdot\pmb{v}_{j}-2v_{j}^2 +\frac{3}{2}(\pmb{v}_{j}\cdot\pmb{n}_{ij})^2\bigg]+\frac{1}{c^2} {\sum_{j\neq{i}}}\frac{G m_j}{r_{ij}^2}\pmb{n}_{ij}\cdot(4\pmb{v}_{i}-3\pmb{v}_j)(\pmb{v}_i-\pmb{v}_j),
\end{equation}

\begin{align}
\pmb{a}_{\rm 2PN} =& \frac{1}{c^4}\sum_{j\neq i}\frac{Gm_j\pmb{n}_{ij}}{r_{ij}^2}\bigg[-2v_j^4 + 4v_j^2(\pmb{v}_i\cdot\pmb{v}_j)-2(\pmb{v}_i\cdot\pmb{v}_j)^2+\frac{3}{2}v_i^2(\pmb{n}_{ij}\cdot\pmb{v}_j)^2 + \frac{9}{2}v_{j}^2(\pmb{n}_{ij}\cdot\pmb{v}_j)^2-6(\pmb{v}_i\cdot\pmb{v}_j)(\pmb{n}_{ij}\cdot\pmb{v}_j)^2-\frac{15}{8}(\pmb{n}_{ij}\cdot\pmb{v}_j)^4 \notag \\
&-\frac{57}{4}\frac{G^2m_i^2}{r_{ij}^2}-9\frac{G^2m_j^2}{r_{ij}^2}-\frac{69}{2}\frac{G^2m_im_j}{r_{ij}^2}+\frac{Gm_i}{r_{ij}}
\Big(-\frac{15}{4}v_i^2+\frac{5}{4}v_j^2-\frac{5}{2}(\pmb{v}_i\cdot\pmb{v}_j) +\frac{39}{2}(\pmb{n}_{ij}\cdot\pmb{v}_i)^2-39(\pmb{n}_{ij}\cdot\pmb{v}_i)(\pmb{n}_{ij}\cdot\pmb{v}_j)+\frac{17}{2}(\pmb{n}_{ij}\cdot\pmb{v}_j)^2\Big)\notag \\
&+\frac{Gm_j}{r_{ij}}\Big(4v_j^2-8(\pmb{v}_i\cdot\pmb{v}_j)+2(\pmb{n}_{ij}\cdot\pmb{v}_i)^2-4(\pmb{n}_{ij}\cdot\pmb{v}_i)(\pmb{n}_{ij}\cdot\pmb{v}_j) -6(\pmb{n}_{ij}\cdot\pmb{v}_j)^2\Big)\bigg]\notag\\
+&\frac{1}{c^4}\sum_{j\neq i}\frac{Gm_j\pmb{v}_{ij}}{r_{ij}^2}\bigg[\frac{Gm_i}{r_{ij}}\bigg(\frac{55}{4}(\pmb{n}_{ij}\cdot\pmb{v}_j)-\frac{63}{4}(\pmb{n}_{ij}\cdot\pmb{v}_i)\bigg) -2\frac{Gm_j}{r_{ij}}\bigg((\pmb{n}_{ij}\cdot\pmb{v}_i)+(\pmb{n}_{ij}\cdot\pmb{v}_j)\bigg)+v_i^2(\pmb{n}_{ij}\cdot\pmb{v}_j)+4v_j^2(\pmb{n}_{ij}\cdot\pmb{v}_i)\notag \\
&-5v_j^2(\pmb{n}_{ij}\cdot\pmb{v}_j)-4(\pmb{v}_{i}\cdot\pmb{v}_j)(\pmb{n}_{ij}\cdot\pmb{v}_{ij}) -6(\pmb{n}_{ij}\cdot\pmb{v}_i)(\pmb{n}_{ij}\cdot\pmb{v}_j)^2+\frac{9}{2}(\pmb{n}_{ij}\cdot\pmb{v}_j)^3
\bigg],
\end{align}

\begin{equation}\label{eq:a3b1pn}
	\pmb{a}_{\rm 1PNc} = \frac{1}{c^2}\sum_{j\neq i}\frac{Gm_{j}\pmb{n}_{ij}}{r_{ij}^2} \left[{\sum_{k\neq{i,j}}}\left(\frac{G m_k}{r_{jk}}+4\frac{G m_k}{r_{ik}}-\frac{G m_{k}r_{ij}}{2r_{jk}^2}(\pmb{n}_{ij}\cdot\pmb{n}_{jk})\right)\right]-\frac{7}{2c^2} {\sum_{j\neq{i}}}\frac{G m_j}{r_{ij}} {\sum_{k\neq{i,j}}}\frac{G m_{k}\pmb{n}_{jk}}{r_{jk}^2},
\end{equation}

\begin{equation}
	\pmb{a}_{\rm 2.5PN} = \frac{1}{c^5}\sum_{j\neq i}\frac{4G^2m_im_j}{5r_{ij}^3}\bigg[\pmb{n}_{ij}(\pmb{n}_{ij}\cdot\pmb{v}_{ij})\bigg(\frac{52}{3}\frac{Gm_j}{r_{ij}}-6\frac{Gm_i}{r_{ij}}+3v_{ij}^2\bigg) +\pmb{v}_{ij}\bigg(2\frac{Gm_i}{r_{ij}}-8\frac{Gm_j}{r_{ij}}-v_{ij}^2\bigg)
\bigg],
\end{equation}
where $G$ is the Newtonian constant, and $c$ is the speed of light, and $r_{ij}=\left|\pmb{r}_i-\pmb{r}_j\right|$, $\pmb{n}_{ij}=(\pmb{r}_i-\pmb{r}_j)/r_{ij}$, and $\pmb{v}_{ij}=\pmb{v}_i-\pmb{v}_j$. Here $\pmb{r}_{i,j}$ are position vectors of body $i$ and $j$, and $\pmb{v}_{i,j}$ are velocity vectors respectively.


\bsp	
\label{lastpage}
\end{document}